\documentclass[journal]{IEEEtran}

\ifCLASSINFOpdf
\else
   \usepackage[dvips]{graphicx}
\fi
\usepackage{url}

\usepackage{graphicx}
\usepackage[]{graphicx}
\usepackage{xcolor}

\graphicspath{{./figs/}}
\DeclareGraphicsExtensions{.eps,.pdf,.jpeg,.png,.jpg}

\begin{document}

\title{VeHIF: An Accessible Vegetation High-Impedance Fault Data Set Format}

\author
{Douglas P. S. Gomes, Cagil Ozansoy
}

\markboth{PREPRINT}
{Gomes and Ozansoy \MakeLowercase{\textit{et al.}}}
\maketitle

\begin{abstract}
High-impedance faults are a challenging problem in power distribution systems. They often do not trigger protection devices and can result in serious hazards such as igniting fires when in contact with vegetation. The current research field dedicated to studying these faults is extensive but suffers from a constraining bottleneck of a lack of real experimental data. Many works set to detect and localize such faults rely on high-impedance fault low-fidelity models, and the lack of public data sets makes it impractical to have objective performance benchmarks. This letter describes and proposes a format for a data set of more than 900 vegetation high-impedance faults funded by the Victorian Government in Australia recorded in high-sampling resolution. The original data set is public, but it was made available through an obscure format that limits its accessibility. The presented format in this letter uses the standard hierarchical data format (HDF5), which makes it easily accessible in many languages such as MATLAB, Python, C++, and more. The data set compiler and visualizer script are also provided in the work repository \footnote{\url{https://github.com/dougpsg/hif_vegetation_data}}.
\end{abstract}


\IEEEpeerreviewmaketitle

\section{Introduction}

\IEEEPARstart{V}{egetation} high-impedance faults (VHIFs) in power distribution systems are characterized by the unintentional current flow (fault) from electrical equipment and nearby vegetation where the resulting current is lower than protection devices' sensitivity. Despite not representing a threat to power equipment, VHIFs are still dangerous disturbances with significant potential to create safety risks and fire hazards. The research field studying high impedance faults (HIFs) are relatively mature \cite{review}, but despite being responsible for many developments, there are no consensus solution or standards for the detection or addressing the problem. While the term HIF is often used interchangeably with VHIFs, it is fair to say that the latter is a sub-problem of the former, probably deserving separate treatment.

The damages created by the 2009 Black Saturday bushfires \cite{creports2} — a series of devastating bushfires with some ignited by electrical faults — highlighted the importance of addressing these faults. The commission responsible for investigating the fires explicitly recommended further research in the field, resulting in a \$750 million funding round. One of the projects, the `Vegetation Conduction Ignition Testing', set to test around twenty plant species were tested in hundreds of experiments on a real three-wire 22-kV feeder to study fault behaviour, and foster protection technology development \cite{marxsenmain}. The tests resulted in an extensive data set of vegetation HIF fault recordings with high-resolution sampling and wideband signals, having a significantly high potential for insights.

Having a publicly available data set of VHIFs is highly significant to its related research field. It is fair to state that one of the bottlenecks for the progress in the field of HIFs is the lack of shared benchmarks and data sets. Most papers either make use of a private data set or rely on synthetic data sets resulting from HIF model simulations (a large share of works) \cite{review}. To the data set being discussed here (VeHIF), a few works have been published using it \cite{gomes2018,gomes20192,kandanaarachchi2020early}, mainly by this letter's authors. One impediment for accessing the data is that, although public, it was disclosed in an obscure format given by the sampling equipment used. Having an easily accessible version of this data set where others can quickly access and interface in many programming interfaces is therefore much valuable to the field. 

Given the limited accessibility of the data set and to collaborate with future research on the complex and evolving problem of VHIF detection, this letter presents a proposed format for the data set, access to the re-organised data, and example code on how to use it.
The described data set format is compiled from the data provided by the Victorian Government data website \cite{testsdata}, but the proposed format offers seamlessly access for experiments. As the data set contains faulty and non-faulty example signals, this format suits investigations that approach this as a classification problem. The data set is publicly hosted on the data science platform Kaggle \cite{kaggledata}, and repository with codes to the data set compiler and visualizer are also provided \cite{githubrepo}. It is important to note that the data set should be used in the context presented and described by the official tests report \cite{marxsenmain}. There, one can find detailed descriptions of the characteristics of the tests, vegetation species tested, electrical configurations and specifications. 

\section{The vegetation high-impedance fault experiments}

The data was made public by the Victorian Government \cite{testsdata}, and it contains logs, video and signal recordings, graphs, and analysis of the tests. The data set described here only includes tests (waveforms) recordings and some metadata regarding the experiments. Such data should be enough for primary analyses and using it in applications like signal classification and spectrum analysis. 
The voltage and current signals were made available in a proprietary format given by the hardware used in their recordings, thus not directly accessible for manipulation. 

Two sampling channels were used for each current and voltage signals. The project team decided to have a channel dedicated to higher frequencies without the influence of the main power frequency (50 Hz). All four channels were connected to an analogue 6-pole Bessel low-pass filter with 10 MHz corner frequency for anti-aliasing. A second low-pass anti-aliasing filter (Bessel IIR digital) with proper data decimation was added to the low-frequency (LF) and high-frequency (HF) channel; It had a cut-off frequency of 50 kHz in LF channels and 1 MHz in HF channels. HF channels were also filtered by a 10 kHz high-pass filter, characteristic of the capacitive voltage divider design. The resulting effective band of LF channels was approximately 0 Hz to 50 kHz, while the HF channels were effectively composed of frequencies from 10 kHz to 1 MHz. 

The high-speed recorder continuously sampled all channels at 100 kSa/s, plus the additional measurements in the HF channels. The latter was not sampled continuously but by periodic intermittent short recordings (snapshots) referred to as `sweeps' with a duration of 20 ms. The sweeps were triggered by a non-synchronous signal with 1 s period, meaning that HF recordings were composed of one sweep per second (one power cycle per second). The many sweeps per test were extracted and sequentially concatenated in a single vector in the data set. The trigger signals are also included in the data, with their square waves signalling the starting time of a sweep recording. 

\section{VeHIF Data set format}

\subsection{The HDF5 format}

Hierarchical data format (HDF) has many advantages and characteristics that fit the use for this data set. It is free, supported by many commercial and non-commercial software platforms and programming languages (MATLAB, C, C++, Fortran, Java), and has efficient partial I/O access to data on disk. Not surprisingly, it has become of wide use in public benchmark data sets and frameworks in fields like deep learning \cite{van2015blocks}.

The file format is hierarchically represented main by two entities: Groups and Data sets. Data sets, in this case, are the actual data, often organized and accessed as arrays (slicing is also a feature). Groups are the hierarchical representation (nodes) where sub-groups and data sets are mapped. A close analogy is the one given by a file system in a computer where `groups' are directories and `data sets' the files. In addition, groups and data sets can have attributes, which can be seen as the metadata of a given node. Groups, data sets, and attributes are all the concepts necessary to work with the VeHIF data set. 

\subsection{Data set structure}

The root of the data set contains only two groups: `cal' and `test'. The latter are tests that were labelled are background noise recordings by the program. As these are recordings where no vegetation was being tested, they can be used as the negative class when using it as training data in a classification problem. The former, `test', represents all the tests available on the data provided in the official data web page \cite{testsdata}, which did not result in an error when translating the data from the original format to the use presented here. In total, there are 53 background and 983 fault tests, as 15 fault tests produced errors in the data set compilation. Both of these root groups contain one attribute, which is a list of the original test number as per the original report \cite{marxsenmain}.

Every sub-group under the two root group represent a unique test recording. They are labelled by their original number and contains at least four data sets: `current\_hf', `current\_lf', `voltage\_hf', `voltage\_lf'. The tests subgroup contains one additional data set, `hf\_trigger', which represents the waveform used to trigger the high-frequency sweeps recordings. The sub-groups under `cals' have two attributes: `filename' and `cal\_type'; the first refers to the original filename and the second to the electrical configuration of the test (phase-to-earth or phase-to-phase). The sub-groups under `test' have four attributes: `filename', `test\_type', `max\_current', and `report\_validity'. Test type can be `bush', `grass', `phase-to-earth', or `phase-to-phase'. The maximum current and report validity are extracted from the program's report. Each test was set to end at a defined current value, and some tests were labelled as invalid as per the test report \cite{marxsenmain}.

\begin{figure}[t]
\centering
\includegraphics[scale=0.4]{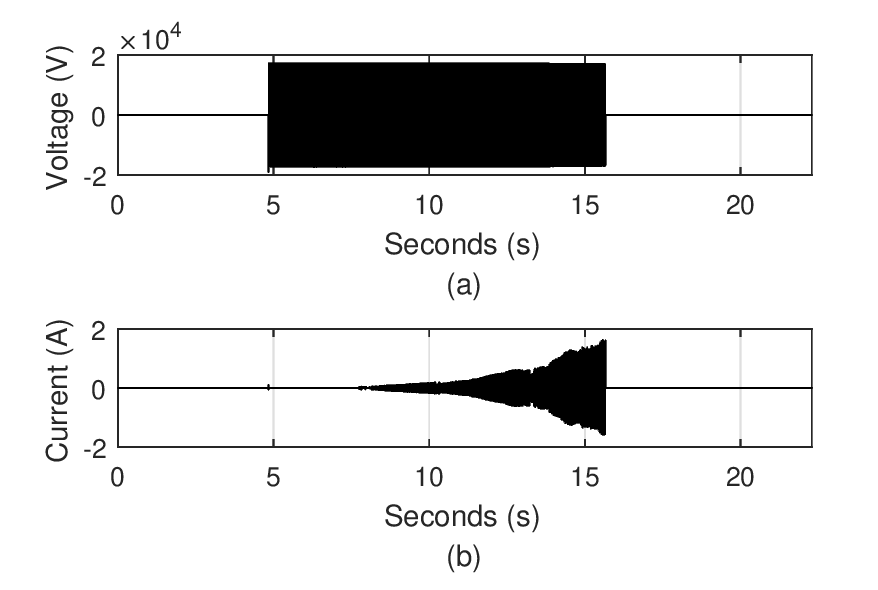}
\caption{Voltage (a) and current (b) waveforms sampled from the LF channel. One of the few tests with pre-fault voltage recordings.}
\label{lf_example}
\end{figure}
\begin{figure}[t]
\centering
\includegraphics[scale=0.6]{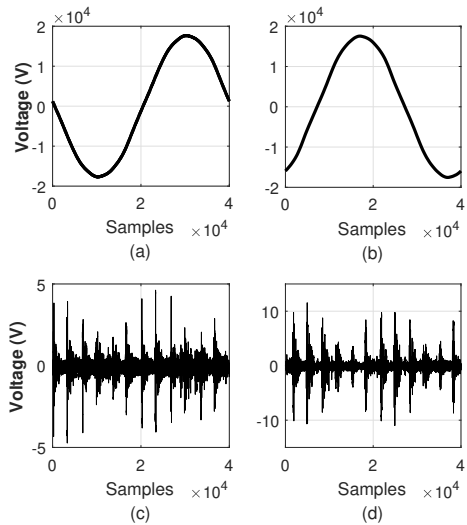}
\caption{Two example of sampled sweeps. LF recordings from a) Non-fault sweep and b) Faulty sweep. HF recordings from c) Non-fault sweep and d) Faulty sweep.}
\label{two_tests}
\end{figure}

Two examples are illustrated in the presented figures. Figure \ref{lf_example} shows the recordings of voltage and current sampled from the LF channel. Differently from the HF channel, this channel samples continuously, and thus it allows to plot the waveform against time throughout the whole duration of the test. Figure \ref{two_tests} illustrates two sweeps recordings from the HF channel. This channel sampled the tests in sweeps of 20 ms with 2 MSa/s sampling rate, leading to 40k samples per sweep (axis in the figure). The HF sampling channel sampled signals connected to the low-pass and high-pass filters, leading to LF and HF pairs of the same sweeps.  Figure \ref{two_tests} shows the pairs (LF and HF) of voltage waveforms for the case where a fault is present (faulty) and in the absence of fault (non-fault). Note how the small currents from such vegetation faults do not introduce any visible effect on the voltage waveforms.

\subsection{Accessibility}

The data set is publicly hosted on the data science platform Kaggle \cite{kaggledata}. In addition to providing the data set in the accessible H5DF format, the data set is accompanied by a repository with the code used to compile the data set and a visualizer script in MATLAB \cite{githubrepo}. As previously mentioned, such tests were thoroughly described in the official report, and one should refer to it for specific information on the tests and experimental configurations. Given the public libraries such as the `h5py' in Python and native H5DF integration to MATLAB, access to the tests waveforms should be seamlessly and direct. 

\subsection{Potential uses}

While facilitating access to the data, such a data set format also allows for quicker experimentation. The proposed categorisation in groups suits investigations approaching the problem in a supervised learning fashion. The group can represent classes, allowing for an objective comparison of methods of studying them. The large number of waveforms can also be used in investigations concerning signal features or patterns. As an example, using machine learning, as in unsupervised learning, or dimensionality reduction approaches to find relevant patterns in the signal. The VeHIF problem still presents significant knowledge gaps, with the behaviour still to be clearly defined. Researchers can benefit from such a format to enhance or objectively compare distinguished investigations.

\bibliographystyle{IEEEbib}
\bibliography{references}

\end{document}